# HTSC Cuprate Phase Diagram Using a Modified Boson-Fermion-Gossamer Model Describing Competing Orders, a Quantum Critical Point and Possible Resonance Complex

Richard H. Squire [ξ], Norman H. March*, M. L. Booth [ξ]

[ξ] Department of Chemistry, West Virginia University
Institute of Technology University
Montgomery, WV 25303, U

* Department of Physics, University of Antwerp (RUCA), Groenborgerlaan 171,
B-2020 Antwerp, Belgium
and
Oxford University, Oxford, England

ABSTRACT

There has been considerable effort expended towards understanding high temperature superconductors (HTSC), and more specifically the cuprate phase diagram as a function of doping level. Yet, the only agreement seems to be that HTSC is an example of a strongly correlated material where Coulomb repulsion plays a major role. This manuscript proposes a model based on a Feshbach resonance pairing mechanism and competing orders. An initial BCS-type superconductivity at high doping is suppressed in the two particle channel by a localized preformed pair (PP) [1] (circular density wave) creating a quantum critical point (QCP). As doping continues to diminish, the PP then participates in a Feshbach resonance complex that creates a new electron (hole) pair that delocalizes and constitutes HTSC and the characteristic dome [2]. The resonant nature of the new pair contributes to its short coherence length. The model we propose also suggests an explanation (and necessity) for an experimentally observed correlated lattice that could restrict energy dissipation to enable the resonant Cooper pair to move over several correlation lengths, or essentially free. The PP density wave is responsible for the pseudogap as it appears as a "localized superconductor" since its density of states and quasiparticle spectrum are similar to those of a superconductor (Peierls-Frolich theory), but with no phase coherence between the PP.

**1. Introduction.** Current studies in high temperature superconducting cuprates (HTSC) have fascinating connections with a number of theoretical issues, many of which have no consensus yet, such as conflicted Hamiltonians with competing multiple orders, the origin of the pseudogap and its relation to both superconductivity and antiferromagnetic Mott insulators, the pairing mechanism, the existence (or not) of a quantum critical point, the role of BEC vs. BCS, and the universality (or not) of fundamental physical parameters to name a few. One of us began our approach to HTSC before some of these were recognized as issues by examining crossover work by Eagles [3] as a function of carrier density.

Egorov and March [4] studied carriers in high $T_c$ cuprates moving through an assembly of Cu ions exhibiting antiferromagnetic spin fluctuations. Employing the results of two-dimensional Fermi liquid theory for this system [5], the in-plane resistivity R is related to the magnetic susceptibility $\chi(\underset{\sim}{Q})$ at the antiferromagnetic wave vector $\underset{\sim}{Q}$ by

$$R \propto T^2 \chi(\underset{\sim}{Q}) \qquad (1)$$

If $\chi(\underset{\sim}{Q})$ has Curie-Weiss behavior ($\propto 1/T$), with $T$ the absolute temperature, then one recovers the characteristic $R \propto T$ form. Combining eq. (1) with the corresponding result

$$(TT_1)^{-1} \propto \chi(\underset{\sim}{Q}) \qquad (2)$$

for the temperature dependence of the nuclear spin lattice relaxation time $T_1$, they eliminated the susceptibility $\chi$ between eqs. (1 and 2) to find

$$RT_1 \propto T \qquad (3)$$

from the two dimensional Fermi liquid assumption. Their use of experimental data for R and $T_1$ led to the result shown in [4, Fig 2]. Eq (3), while reproduced well above $T_c$, deviates from the experimental curve as $T_c$ is approached from above. This was interpreted as an **experimental fingerprint of the formation of precursor 2e bosons** [6, 7]. The evidence for such pairing above $T_c$ in the cuprates has been subsequently reviewed by Timusk and Statt [8].

Spurred on by the concept of real space 2e bosons / Cooper pairs, we reexamined other work such as Leggett's [9] variationally calculated crossover at T = 0 where he obtained a dilute, attractive Fermion gas flowing in a continuous crossover from large, overlapping BCS Cooper pairs to a tightly bound condensate of diatomic molecules. Later, Nozieres and Schmitt-Rink (NSR) [1] extended Leggett's work to finite temperatures and lattice models. Using a factorizable potential, they found fermions formed local singlet pairs and behaved as hard core bosons which have been named preformed pairs (PP). Later, we will show that these pairs become the pseudogap which are phase disordered, local pairs of electrons. NSR also examined the free energy of pairs and the effects on a superfluid phase transition as BCS theory crosses over to BEC; the transition temperature $T_c$ smoothly evolves as a function of attractive coupling a does the quasi-particle spectrum. While the two limits were well established, interpolation was necessary in between, the location of Cuprates and fulleride HTSC.

Based on NSR, we presented an electronic phase diagram for fullerides using a modified Boson-Fermion Model (MBFM) [2, 10] where the boson function in the theory is carried out by two doped electrons onto a fullerides molecule. These two electrons form a singlet state which we have proposed is a localized charge 2 density wave. It seems as though this density wave can be experimentally detected as localized (thereby insulating) superconductivity since an energy gap is formed (appendix A). This result suggests an explanation for a "Cooper pair" insulator, also recently found in other systems [11].

Can such a localized pairing be applicable in a cuprate HTSC? Recently, there have been a number of theories brought into question by new experimental results such as polarized neutron diffraction spectroscopy [12]. Indeed, some may have already been eliminated by interesting STM [13] and ARPES [14] results that to our knowledge have not been satisfactorily explained by any one theoretical model. In order to more fully explain a phase diagram with competing orders, the MBFM model needed to be combined with the so-called Gossamer model [15] which deals with competing orders. Then, the combined model suggests possible explanations for many features of the phase diagram (such as the shape of the SC dome with a QCP hidden beneath it, a checkerboard pattern observed in STM, a resonance-type superconductivity, and a d wave Cooper pair wave function overlapping with an experimentally observed Fourier transformed spin correlation function which leads us to suggest a resonance "complex" is responsible for pairing, all defined later.

The outline of what follows concerns first in Section 2 a discussion of the merger of the MBFM with the Gossamer model, thereby providing an updated viewpoint of the relevant interacting orders including quasiparticle theory. Section 3 is an assessment of the present understanding of fulleride superconductivity in the framework of the model with emphasis on the formation of a 2e-DW. Then, Section 4 covers model application to cuprate HTSC with comparison experimental results (which really generated the theory), plus some proposals. Finally, Section 5 contains of a summary plus some suggestions for future work that might prove fruitful.

**2. Modified Boson-Fermion-Gossamer Model.**

**a. Modified Boson-Fermion Model**
Our initial work on fullerides suggested a fullerides molecule with two doped electrons formed a charge 2 density wave isolated on a single molecule. This density wave has aspects of an isolated Cooper pair (see Section 2b). Since we were interested in what NSR might describe as a hard core boson or electron-molecule [1, 9] (two confined, highly correlated electrons) interacting with a third electron, it seemed only natural to examine cold atom work where this type of interaction occurs [16]. In the simplest description because this field has truly blossomed, a host of say, Rb atoms, all in the same quantum state are cooled to picokelvin temperatures; then, some atoms are detuned to avoid the Pauli principle and bosons are formed from two atoms by passing through a Feshbach resonance [17, 18] with the application of a variable magnetic field. One model which can describe this type of interaction is the boson fermion model (BFM), created by Ranninger and collaborators [19]. This model was also studied by Friedberg and Lee [20] since the experimentally measured HTSC short coherence length was speculated to be similar to resonance in particle work. Schafroth [21], and Leggett [9] had initiated the original argument for real space as opposed to momentum space pairing, as exists in the BCS model. The authors above who use the BFM set the BCS interaction U = 0.

We have designated the modified BFM (MBFM) to specify $U_{BCS} \neq 0$. A direct comparison and insight can be made with cold superfluid Fermi gases (following Ohashi and Griffin [22, esp. 22c], and Chen et al [23], and others [17, 18] and keeping only the essential terms for the non-resonant and resonant Cooper pair / Boson molecule interaction (see [2a], Appendix C for the full Hamiltonian):

$$H = \sum_{\bar{p},\sigma} \varepsilon_{\bar{p}} c^{\dagger}_{\bar{p}\sigma} c_{\bar{p}\sigma} - U \sum_{\bar{p},\bar{p}'} c^{\dagger}_{\bar{p}+\bar{q}/2,\uparrow} c^{\dagger}_{-\bar{p}+\bar{q}/2,\downarrow} c_{-\bar{p}'+\bar{q}/2,\downarrow} c_{\bar{p}'+\bar{q}/2,\uparrow} \quad (4)$$
$$+ \sum_{q} \left( E^{0}_{\bar{q}} + 2\nu \right) b^{\dagger}_{\bar{q}} b_{\bar{q}} + g_{r} \sum_{\bar{p},\bar{q}} [b^{\dagger}_{\bar{q}} c_{-\bar{p}+\bar{q}/2,\downarrow} c_{\bar{p}+\bar{q}/2,\uparrow} + h.c.]$$

Here $c_{\bar{p}\sigma}$ and $b_{\bar{q}}$ represent the annihilation operators of a Fermion (Fermi atom) with kinetic energy $\varepsilon_{\bar{p}} = p^2/2m$ and a quasi-molecular Boson with the energy spectrum $E^{0}_{\bar{q}} + 2\nu = q^2/2M + 2\nu$, respectively.

In the second term including 2e DW-BCS interaction ($-U < 0$ is the BCS theory) provides an attractive interaction from non-resonant processes. The threshold energy of the composite Bose particle energy band is denoted by $2\nu$ in the third term. The last term is the Feshbach resonance (FR, coupling constant $g_r$) that describes how a b-Boson (again, an electron "molecule" in "cold" atom parlance) can dissociate into two Fermions, or how two Fermions can bind into a b-Boson. Since the b-Boson "molecule" is constructed from a bound state consisting of two Fermions, the boson mass is $M = 2m$ and the conservation of total number of particles N imposes the following number relationship

$$N = \sum_{\bar{p}\sigma} \left\langle c^{\dagger}_{\bar{p}\sigma} c_{\bar{p}\sigma} \right\rangle + 2 \sum_{\bar{q}} \left\langle b^{\dagger}_{\bar{q}} b_{\bar{q}} \right\rangle \equiv N_F + N_B \quad (5)$$

Incorporating this constraint into eq (4) again results in a "grand canonical Hamiltonian", as used previously for variable particle number, since b bosons ("molecules") are formed from fermions, and vice-versa. With this relationship, there is only one chemical potential,

$$H - \mu N = H - \mu N_F - 2\mu N_B$$

and it leads to an energy shift

$$\varepsilon_{\bar{p}} \rightarrow \xi_{\bar{p}} \equiv \varepsilon_{\bar{p}} - \mu$$

and

$$\varepsilon_{B\bar{q}} + 2\nu \rightarrow \xi_{B\bar{q}} \equiv \varepsilon_{B\bar{q}} + 2\nu - 2\mu$$

From this point we just present most of the essential features necessary for a superconductor phase diagram (see [2] for details). When the particle-particle vertex develops a pole at $\vec{q} = v_n = 0$, a superfluid phase transition occurs [24] which corresponds to the following "gap" equation for $T_c$,

$$1 = \left(U + g_r^2 \frac{1}{2v - 2\mu}\right) \sum_{\vec{p}} \frac{\tanh\left(\varepsilon_{\vec{p}} - \mu\right)/2T_c}{2\varepsilon_{\vec{p}} - 2\mu} \qquad (6)$$

In eq (6) $g_r^2/(2v - 2\mu)$ is the additional pairing interaction mediated by a boson which becomes very large when $2\mu \to 2v$. $T_c$ is the temperature at which instability occurs in the normal phase of a degenerate Fermi gas due to formation of bound states with zero center of mass momentum $(\vec{q} = 0)$ and energy $2\mu$. The number and gap equations are thus appropriately modified versions of the BCS gap equation. Schrieffer's singlet pair wave function,

$$\Psi(\vec{r}_2, \vec{r}_1) = \langle N-1 | \psi_\downarrow(\vec{r}_2) \psi_\uparrow(\vec{r}_1) | N \rangle \qquad (7)$$

where N represents the number of pairs, and $\psi_\downarrow(\vec{r}_1)\psi_\uparrow(\vec{r}_2)$ annihilates a pair. Because the electrons are created in pairs, the effects of anticommutation of electrons (Pauli principle) are negated, so this function takes a macroscopic value in a fermion system for an idealized condensed pair pairs. For large N we can consider

$$\Psi(\vec{r}) = \langle N-1 | \psi(\vec{r}) | N \rangle \qquad (8)$$

as an order parameter [25] which can then be used in the Landau-Ginsburg free energy equation contained in the Gossamer analysis below. $\Psi(\vec{r})$ is suitable for the entire range from the BCS limit of large, overlapping pairs to the Bose limit of small, essentially non-overlapping pairs. The quasi particle spectrum was previously described [10, fig x therein], and remains "BCS-like" with certain gap modifications [26, 27].

So, applying MBFM theory to a degenerate Fermi gas with a strong pairing interaction results in a strong suppression of $T_{BCS}$ caused by fluctuations in the two-particle channel (which results in a 2e DW). In the strong coupling regime there are two "types" of Bosons even above $T_c$, "molecules" ("molecules" in "cold" atom parlance) associated with the resonance and pre-formed pairs, as previously discussed [1, 2, 22; also see Fig 1a]. As we interpret the model, the pathway from BCS to BEC is a phase transition.

Even though the two body scattering length changes abruptly at the unitary scattering condition, there is a less dramatic change in the nature of the superconductivity and certain physical properties as the SC shifts from BCS-like to FR superconductivity [28, 29]. As the unitary limit is traversed going from high doping to low, the BCS Cooper pairs are suppressed as the location of a critical (unitary) point in the fulleride phase diagram is passed. It seems well established in high-$T_c$ studies that properties are different on one side of the SC "dome" relative to the other. In addition some features of the evidence for a QCP such as quasiparticle lifetime variation have been present [30, 31, 32]. Using the connection we have established with the cold atom work, it seems obvious that at T = 0 the Feshbach resonance should dominates as it is electronic in nature, whereas other modes usually have the energy depleted. One of the features with the QCP here is presumably the order parameter influence of the 2e DW on both the BCS SC and the Feshbach resonance. It seems appropriate to suggest that this interaction even thought it is in the two particle channel, could result in a QPT [31, 32] from a BCS superconductor to a BFM superconductor; indeed, the Gossamer model considers a metal to metal transition [15], so a SC to SC transition at lower temperature seems plusible. Thus, fullerides, as a class of compounds could have at least four nearby phases influenced by doping – insulator, metal, BCS and FR SC, and possibly a fifth, preformed pairs in a metal phase causing a "boson" metal. Further support for the notion of a QPT is obtained in a recent article describing QPT's in the interacting boson model (IBM) in the nuclear case [33].

**b. Gossamer model [15]**

We turn to a mean field transition treatment of a charge density wave where the lattice distortion of collective renormalized vibrations can be shown to be (see Appendix I for more details)

$$\langle u(x) \rangle = \Delta u \cos(2k_F x + \phi) \qquad (9)$$

Opening an energy gap at a transition temperature, $T_{CDW}^{MF}$,

$$k_B T_{CDW}^{MF} = 1.14 \varepsilon_0 e^{-1/\lambda} \qquad (10)$$

with $\lambda$, the electron-vibration coupling constant (dimensionless) and a total (gap) energy

$$\Delta = 2\varepsilon_F e^{-1/\lambda} \qquad (11)$$

and condensation energy (relative to the normal state) of

$$E_{cond} = E_{norm} - E_{CDW} = \frac{n(\varepsilon_F)}{2}(\Delta^{CDW})^2 \qquad (12)$$

Note that eq (10) is exactly like the BCS equation; also see eq (A3), Appendix A. In a quasi one-dimensional system containing superconductivity and CDW studied by Levin et al [34], Bilbro and MacMillan [35], and Balsiero and Falicov [36], the two instabilities are incompatible, as is now well known. Thus, a typical CDW state tends to suppress BCS superconductivity; a 2e DW does also due to strong fluctuations in the particle-particle channel. Machida et al [37] have studied the effect of the CDW state on superconductivity as well as the opposite case; there is a temperature domain where both states could coexist. Later we will see that a charge 2 density also competes with BCS SC.

In fullerides the competition between CDW and BCS superconductivity is an s-wave competition that is exact. An order parameter for a CDW and s-wave superconductor (SSC) may be written as, respectively,

$$i\Delta_{CDW} f(\vec{k}) = \left\langle \sum_\sigma c^\dagger_{\vec{k}\sigma} c_{\vec{k}+\vec{Q}',\sigma} \right\rangle \text{ while } \Delta_{SSC} f(\vec{k}) = \left\langle c_{\vec{k}\uparrow}, c_{-\vec{k}\downarrow} \right\rangle \qquad (13)$$

Following [26, 38], when U > 0 the ground state is an ordered antiferromagnet with expectation values

$$\begin{bmatrix} \langle S_j^x \rangle \\ \langle S_j^y \rangle \\ \langle S_j^z \rangle \end{bmatrix} = \frac{1}{2} \begin{bmatrix} \langle c_{j\uparrow}^\dagger c_{j\downarrow} + c_{j\downarrow}^\dagger c_{j\uparrow} \rangle \\ i \langle c_{j\uparrow}^\dagger c_{j\downarrow} - c_{j\downarrow}^\dagger c_{j\uparrow} \rangle \\ \langle c_{j\uparrow}^\dagger c_{j\uparrow} - c_{j\downarrow}^\dagger c_{j\downarrow} \rangle \end{bmatrix} \quad (14)$$

On the other hand, when U < 0, the ground state is a degenerate mixture of density wave checkerboard order and s-wave superconductivity with expectation values

$$\begin{bmatrix} \langle \mathrm{Re}(\Delta_j) \rangle \\ \langle \mathrm{Im}(\Delta_j) \rangle \\ \langle n_j \rangle \end{bmatrix} = \begin{bmatrix} \langle c_{j\uparrow}^\dagger c_{j\downarrow} + c_{j\downarrow}^\dagger c_{j\uparrow} \rangle \\ i \langle c_{j\uparrow}^\dagger c_{j\downarrow} - c_{j\downarrow}^\dagger c_{j\uparrow} \rangle \\ (-i)^j \langle c_{j\uparrow}^\dagger c_{j\uparrow} + c_{j\downarrow}^\dagger c_{j\downarrow} \rangle \end{bmatrix} \quad (15)$$

The two orders are energetically degenerate, occur simultaneously, and are related by a rotation, hence this type of order interaction could result a quantum critical point. The quasiparticle energy dispersion becomes

$$E_k = \pm \sqrt{\left[ \left( \varepsilon_{\vec{k}}^2 + |\Delta^{CDW}|^2 \right)^{1/2} \pm \mu \right]^2 + |\Delta^{SSC}|^2} \quad (16)$$

where $\Delta^{CDW}$ is the CDW gap and $\Delta^{SSC}$ represents the s wave SC gap if $\Delta^{SSC} = zU$ and $\Delta^{CDW} = yU$, we have the condition that the SC order parameter can be rotated into the checkerboard charge order without closing the quasiparticle gap as an exact symmetry. If we write a generic T = 0 Ginzburg-Landau free energy as

$$F = \left( y^2 + |z|^2 \right)^2 + \gamma y^2 |z|^2 - \alpha y^2 - \alpha' |z|^2, \quad (17)$$

Then, eq (17), derived by Chakravarty et al [38], leads to a similar curve (Fig 1a), but for a fundamentally different reason. The competition between a CDW and SC in an s-wave case is exact. To summarize the Gossamer model describes competing order parameters and postulates a strong pair potential which we propose it the Feshbach resonance assisted by charge 2 density waves.

## 3. Fulleride Superconductivity phase diagram (2a).

Starting from the high-doped BCS side of the SC dome, when the threshold $2\nu$ is much larger than $\varepsilon_F$, the Fermi states are dominant. A BCS-like phase transition is present since $\mu \sim \varepsilon_F > 0$ and stable bosons are absent. The only exception is if $2\tilde{\nu} = 2\mu$, then, a stable Boson appears and this condition reduces to the BCS gap equation for $T_c$. This will result in the formation of stable Cooper pair Bosons and an immediate phase transition at the same temperature, namely the BCS theory. In this limit no stable, long-lived Bosons with $\vec{q} \neq 0$ exist above this transition temperature. As doping decreases to, say, somewhere around n = 3.5 electrons, on the way towards the apex in the phase diagram, the unitary point $(\mu = 0)$ is encountered. Electron "molecules" (in the sense of Leggett [9]) with a finite lifetime begin to be formed when $2\nu \leq 2\varepsilon_F$. This is the 2e DW which performs two functions in BCS-BFM superconductivity. The 2e DW order parameter competes with the BCS order parameter. So, when the fulleride doping decreases to about 3.5, the 2e DW begins to dominate BCS superconductivity. The second function of the 2e-DW is to provide a crucial component of the incipient Feshbach resonance. Then, the dome diminishes as doping decreases moving the 2e DW out of resonance.

The pseuodogap origin seems apparent [8, 23]; it is due to the localized pairing (2e DW) without long range order (LRO), as the phases of the DW wave functions of adjacent fullerides are random. As mentioned, the localized 2e density wave has the experimental appearance of a "localized superconductor" because its energy gap equation is the same as the BCS gap equation. In addition its density of states and quasiparticle spectrum are also similar. NMR work by Brouet et al has confirmed the existence of the singlet state [39].

As doping of two electrons is approached, a fulleride molecule maintains a 2e DW (c-type Bosons). The resulting fulleride state has Mott-Hubbard-Wigner crystal-like correlations and the ordered state is accompanied by the appearance of the 2e DW insulating gap. We have proposed that both the fulleride and cuprates superconducting mechanisms are similar; we will illustrate how this is possible below. Certainly a major difference is that fullerides are electron-doped superconductors; however as suggested in a recent review by Lee, Nagosa, and Wen [40], it seems that complete details of the electron interaction we will discuss may well be hidden by strong antiferromagnetism, just as in the cuprates electron-doped SC's (fig 2b). It has also been suggested that holes may be responsible for the superconducting in these electron-doped materials due to charge segregation [41]. Nonetheless, while the fulleride proposal may be of lesser interest for some, the benefit from an understanding of a simpler, s-wave superconducting case may allow a better understanding of the fundamental physical nature of most of the electronic phase diagram of the cuprates (and fullerides). The above proposal also suggests why the metal-insulator (M-I) transition that certain doped fullerides span happens; changes such as pressure in a fulleride structure can move seemingly closely related fullerides from one side of the transition to the other. This effect has recently been observed in cesium fulleride [42].

**4. Mechanism of HTSC in the Cuprates (Fig 2b).**

**a. Pertinent Chemistry and Physical Properties of the Cuprates.**
Excellent reviews of the chemistry and physics of transition metals and the MI transition have been made by Lee et al [40] and Goodenough [43], so we will limit our discussion to very generic topics that could apply to most cuprates, especially focusing on considerations that might suggest a structure analogous to 2e DW formation in fullerides. The perovskite structures were selected by Bednorz and Muller [44] because they were Jahn-Teller active. They picked the correct class to study as superconductivity has not been found in other perovskite-related systems to our knowledge. The transition metal oxides are very appealing in their own right since $U_n$, the intra-atomic electron coulomb energies that separate a $d^n$ from a $d^{n+1}$ configuration are low enough that at least one of these energies lies between the bottom of the 5d or empty s conduction band and the top of the $O^{2-}:2p$ bands. Thus, it is always possible to reduce and/or oxidize a transition metal perovskite $MO_3$ array and create a mixed valence electronic conduction. The copper oxide superconductors are all mixed-valent compounds. There is increasing evidence that an unusual transition accompanied by cooperative oxygen displacements might create a spinodal phase segregation [45], a "charge bag", a charge 2 density wave (2e DW), or some other heterogeneous object at small length scales in a number of transition metal oxides, as evidence by STM spectroscopy [46]. Stabilization of a phase between electrons itinerant and localized is thought to occur and as experiment and theory (MBFM) suggest, these fluctuations might be associated with elementary excitations, thermodynamically consistent with the existence of a quantum critical point (QCP). There are energetically similar order parameters and interesting physical changes, one of which is the orthorhombic to tetragonal phase transition which produces a 40% drop in Young's modulus. The copper-oxygen-copper bond angle distorts from 180° below this transition, thereby permitting increased electron-lattice interactions.

The overall situation has generally been described by a Hubbard model, Fig 1b, where b is a continuous variable running from the Mott-Hubbard insulator through the M-I transition  If there are small interatomic interactions, the on-site correlation energy U is larger that the tight-binding bandwidth W, and $b<b_g$, and interatomic interactions can be described by superexchange perturbation theory leading to a Heisenberg spin-spin Hamiltonian and antiferromagnetic order (fig 3a). As the doping increases, the superexchange perturbation expansion loses its validity since, as $b_g$ is approached, U is lowered and $T_N$ (Neel temperature) decreases. The $b_g$ line represents a semiconductor-metal Mott transition. Interestingly enough, a system with a continuous variation of b through $b_g$ has not been found experimentally including $LaCuO_3$ and $La_2CuO_4$ [43].

The standard view of the metal-insulator (MI) transition for single valent compounds assumes a smooth transition from localized electrons to itinerant ones; an energy gap U is opened in the process as the bandwidth W decreases from W>U to W<U. The metal insulator transition at W ~ U for a half-filled band is known as the Mott-Hubbard transition. Some time ago Brinkman and Rice [47], and March, Swzuki, and Pariello [48] made the prediction that as the transition point is approached from the itinerant electron side that the mass $m_{e-e*}$ becomes infinite at the Mott-Hubbard transition, $U=U_c$, due to on-site electron-electron interactions that increase the bare electron mass, $m_{b*}$. It has often been suggested that previously neglected electron-lattice interactions induce a first order transition to a localized, strongly correlated system [49, 50, 51].

Goodenough has suggested a strong variation of W with mobile hole concentration p introduces an electronic instability in mixed-valent compounds where W ≈ U applies. Since $U>E_p$ ($E_p$ is the pairing energy) typical CDW's are suppressed and dynamic charge fluctuations caused by atomic displacements which we suggest can ultimately result in a 2 electron (or 2 hole) structure, described below, and Fig 3 [52]. This region is appropriate for p-type copper oxides and leads to enhanced electron-phonon interactions, and even though on-site U suppresses mixed valence reactions, the system can achieve stabilization by segregation into carrier-rich and –poor regions in the presence of strong electron-phonon interaction [53]. These interactions are strong enough to change the mean C-O distances and support a cooperative pseudo Jahn-Teller effect. Above $T_c$ there should be anomalous transport, while below will be what Goodenough has suggested is a charge bag similar to Schrieffer's spin bag [54]. Then, there could well be a strong attraction to a second charge bag; the combination we argue might produce a charge 2 bag or charge 2 DW, somewhat similar to the polyacetylene dimer model [55] where two-dimensional charge order leads to a pseudogap. Goodenough and others [10, 20] propose that the electron-phonon coupling is different than BCS as the pair potential is non-retarded. While a compelling argument has been made that HTSC is not a bipolaron condensation [56], further recent work suggests a model that captures many of Goodenough's earlier ideas [57]. Clearly, unusual ordering is taking place; however, this proposal agrees only moderately well with experiment in its detail.

**b. Is the Checkerboard STM Pattern in Cuprates a Manifestation of the above Ideas?**
The history of charge ordering has been discussed for some time in the cuprate superconductors [58]. Here, charge order is most generally defined as an SU(2) invariant observable. Some time ago, static stripes which appeared to co-exist with superconductivity at very low temperatures were found in Nd-doped $La_{2-x}Sr_xCuO_4$ [59]. Theories based on some of those materials where no static order has been observed are close in proximity to a critical point where charge and/or spin order can be established in the superconducting state [60, 61, 62]. The real space period is an even integer of the Cu lattice spacing and is bond-centered [63]. Most of the work in this period focused on spin or magnetic interactions as the pairing in HTSC. The charge density wave was then a result of a spin density wave. This is especially curious as a charge density wave order parameter has been shown to compete with BCS superconductivity [34, 35]; however, this is not the case for d-wave symmetry. Later, t-J model numerical studies at doping $x = 1/8$ have observed a period equal 4 bulk charge ordering in both insulators and superconductors [61, 62, 63]. Then, STM data found an even period nodulation in site charge density and bond kinetic energy in $Bi_2Sr_2CaCu_2O_{8+\delta}$ [13]. However, in a later large-N theory superconductivity order competes with CDW order [64]. A pair density wave has been established in the context of charge ordering near vortices [65, 66, 67]. This type of state was argued to occur naturally in the plaquette approach of Altman and Auerbach; we also used the plaquette suggestion in our early work on fullerides, and indeed, established a relation between the pair density wave and an unusual Abrikosov vortex [68]. The pair wave has also been suggested in the t – J model using large N approximation and both suggest that hole pairing is more stable than two single holes on disconnected holes. From analysis of STM data by Podolsky and others, suggest a simple charge density wave is not consistent with the periodic modulation, but a pairing amplitude is; they call this a generalized density wave [69, 70, 71].

More recently, two general experimental features have been found in HTSC in energy and momentum spaces [72]. They are two entities, dispersive peaks in FT-STM, presumable interference patterns thought caused by scattering from impurities, and a non-dispersive checkerboard order, thought to be a key to comprehending Cuprate SC. There have been several explanations, one of which we describe as a pair density wave (PDW) to distinguish it from fullerides. We consider this structure as an analog to our 2e DW in fullerides. Like the fullerides, it has its origin in a higher energy scale of the Cuprates, suggesting that if this is the pairing responsible for HTSC, it is more easily destroyed by phase fluctuations than pair breaking.

The non-dispersive checkerboard order mentioned above was suggested to be the result of PDW localization. Following Podolsky et al [70] and Seo et al [73], the d-wave checkerboard density (d-WCB) must exist in the d-wave SC when PDW order is present. Further, there is a complementary interaction between particle-hole (PH) and particle-particle (PP) channel orders. Following the s-wave case, eq. (13), an order parameter for a d-DW and d-wave superconductor may be written as, respectively,

$$i\Delta_{dDW} f(\vec{k}) = \left\langle \sum_\sigma c^\dagger_{\vec{k}\sigma} c_{\vec{k}+\vec{Q}',\sigma} \right\rangle \text{ and } \Delta_{dSC} f(\vec{k}) = \left\langle c_{\vec{k}\uparrow}, c_{-\vec{k}\downarrow} \right\rangle \text{ with } \vec{Q}' = (\pi,\pi) \quad (18)$$

A natural PDW order exists in the mixed state of the d-DW and d-SC with wavevector at $\vec{Q}'$ of

$$i\Delta_{dDW}\Delta_{dSC} f^2(\vec{k}) \propto \left\langle c_{\vec{k}\uparrow}, c_{-\vec{k}+\vec{Q}'\downarrow} \right\rangle \quad (19)$$

with the result that the mixed state of d-SC and d-DW can be viewed also as a mixed state of d-SC and PWD. But the symmetries of the orders in the P-P and P-H channels are correlated; if one is a d-wave, the other is an extended s-wave, as shown by replacing $\vec{Q}'$ by $\vec{Q}$. Because of this complementary interaction, orders in both channels need to be considered together if a phase fluctuation leads to a Cooper pair modulation.

Seo et al further verify the complementarity's of the d-WCB and PDW through a self-consistent Bogoliubov-deGennes calculation using a nearest-neighbor density attraction that creates a checkerboard modulation on top of the superconducting state. Their converged results verify the coexistence and symmetry of correspondence between the PDW and d-WCB with the following typical results stated:

$$\Delta_{dsc}(\vec{r}) = \Delta_0$$
$$\Delta_{pdw}(\vec{r}) = \Delta_1 \cos \vec{Q} \cdot \vec{r} \quad (20)$$
$$\Delta_{dwcb}(\vec{r}) = W_0 \cos \vec{Q} \cdot \vec{r}$$

As illustrated in Fig 4, the DWCB has d-wave symmetry and a 4a x 4a periodicity. The symmetries of the DWCB and PDW orders are d-wave and extended s-wave while DWCB and PDW have the same spatial $(\cos \vec{Q} \cdot \vec{r})$ modulation (also similar to eq A4d, Appendix A). These order parameter interactions have been arrived at by others [52c, 70]. It is interesting that Anderson discusses the checkerboard pattern as a Wigner-like crystal [74] which we suggested earlier for the charge 2e DW in doped fullerides [68].

The idea of a charge 2e density wave in the fullerides might have suggested a similar species in the cuprates some time ago if they had been more similar. Using a tabular form of Leggett's recent analysis [30] and with the benefit of hindsight, we propose a modified one, reflecting the fact that these superconductors are more closely related than originally thought. For other exotic superconductors, the final word is certainly not in, but Haas et al [75] has certainly made some interesting comments in this regard. The obvious question to ask is whether the 2e DW are crucial to HTSC; we believe that the answer is absolutely yes.

**4. Experimental Support.**
The comparisons with experimental results seem very favorable and also suggestive.

**a) Heterogeneous structure of copper oxide planes.**
There have been recent scanning tunneling microscope (STM) experiments which have provided tantalizing glimpses of the "CDW" cuprate checkerboard, pseudogap pairing and more. Kohsaka et al [72], reports Cooper pairs that vanish as the Mott insulator is approached in $Ba_2Sr_2CaCu_2O_{8+\delta}$. Here, removing a few electrons from the antiferromagnetic ground state of a cuprates' Mott insulator transforms the system into a superconductor via delocalized Cooper pairs in k-space. When this transformation happens, two types of electronic excitations appear; one, a pseudogap at higher energy, and lower energy Bogoliubov quasiparticles from the breakup of Cooper pairs. Through a novel simultaneous imaging of both r-space and k-space, the Bogoliubov quasiparticle region shrinks rapidly while transferring spectral weight the higher energy r-space states which are not delocalized Cooper pairs, break translational and rotational symmetries at the atomic scale, and are identified as the pseudogap states. These results seem consistent with the combined modified Boson/Fermion/Gossamer model.

**b) HTSC Pair (2e) density Wave.**

Independent STM work also identifies a periodic, oriented four unit cell "checkerboard" pattern surrounding vortex cores in the same material. Using STM Wise et al [76] report a static, non-dispersive, "checkerboard electronic modulation in the cuprates that has strong doping dependence. The description by Seo et al [73] fits this pattern, providing support that an experimentally observed manifestation of the proposed pair density wave is the charge 2e DW which dominates a portion of the electronic phase diagram.

**c) The Cooper pair wave function, checkerboard and possible resonance "complex".**

The real space form of the two dimensional Cooper pair wave function can be approximated as [77],

$$\psi(\vec{r}) \propto \cos(rk_F)(x^2 - y^2)e^{-3r/\xi_0} \qquad (21)$$

Here, $r = \sqrt{x^2 + y^2}$ is the radius, $k_F$ is the Fermi wave vector and $\xi_0$ is the superconducting coherence length, the approximate scale over which $\psi$ is finite. The wave function is presented as a real space entity which is appropriate for a Feshbach resonance (direct) interaction as opposed to a retarded one as in BCS theory. The comparison of the two functions is presented in Fig 5 seems reasonable, since a close interaction has been speculated as one reason that the coherence length is short in the Cuprates [20].

Recent, McDonald et al [78] created a parabolic functional form of neutron scattering data

$$E(\vec{q}) = E_0 \left[ 1 - \left( \frac{a^2 (\vec{q} - \vec{Q}_0)^2}{\pi^2 \delta^2} \right) \right] \tag{22}$$

where $E_0$ is a doping-dependent energy scale and a is the planar lattice parameter. Values for $\delta$ depend on the doping and composition of the cuprate (see [78], Table 1 for values). The largest scattering intensity appears as a cluster of four incommensurate peaks when the value of $E \to 0$ and $\vec{Q} = \left[ \left( \pm \pi/a, \pm (1 \pm 2\delta)\pi/a \right) \right]$ and $\vec{Q} = \left[ \pm \left( (1 \pm 2\delta)\pi/a \right), \pm \pi/a \right]$.

The form is similar to a $d_{x^2+y^2}$ wave function. A spin fluctuation model with a finite correlation length $\xi$ [79, 80] is

$$s(\vec{r}, t) = \sum_{\vec{Q}} s_{\vec{Q}} \exp\left( -\frac{r}{\xi} + i\omega t \right) \cos\left[ \vec{Q}(\vec{r} - \vec{r}_0) \right] \tag{23}$$

with $\omega$, the fluctuation angular frequency, $\vec{r}_0 = (\pm d, 0)$ or $(0, \pm d)$ and $d = a/2\delta = 2/3 \xi_0$ and $\vec{Q}$ assumes the values above. With the proper choice of $\vec{r}_0$ (including a phase difference between adjacent lobes of the spin fluctuation distribution), the Fourier transform of eq (23) represents the neutron scattering data, Fig 7a. Then, noticing the cosine dependence of this function, the Cooper pair wave function and the checkerboard modulation, and specifically the $\cos \vec{Q} \cdot \vec{r}$, there is a rather remarkable mapping overlap of these three entities, Fig 7(a, b, c) at different dopings. The comparisons are very favorable and also suggestive of a resonance "complex" for a Feshbach (direct) interaction.

The linear relationship between d and $\xi_0$ follows from Table I [78] (only in the under doped region). The incommensurate peaks change as doping increases, as has also been observed in STM experiments [76], spreading out as p increases, while the spatial distribution is compressed, as is also the case with the Cooper pair wave function. In addition the d wave DW has a variable nature also, moving to higher energy as doping decreases. It seems quite plausible that the entities are in the best resonance condition at maximum $T_c$, with the resonance decreasing as doping is decreased or increased, as illustrated by arrows in Fig 7.

Other recent experimental work seems to support conclusions reached in this manuscript. [81, 82, 83, 84].

**d. MBFM, Feshbach Resonance Superconductivity, quantum critical point (QCP) and the electronic phase diagram**.

Once the 2e DW is identified as the "modified" preformed pair, the general theory of the MBF-Gossamer Model, developed over the past several years, obtains.

One feature that was not discussed in the original BFM is the occurrence and location of a critical point because the BCS interaction was turned off. As the 2e DW energy approaches $T = 0$, the suppression of the BCS superconductivity is reduced and it is expected that a number of physical properties should dramatically change. The material ought to behave as a typical metal with an intact Fermi surface. Recent experimental evidence suggests this is true [85]. The location of the QCP cannot be positioned by the theory, but presumably, based on the dramatic change in properties, it seems as though it ought to be somewhere between x = 0.20 to 0.24. This is not the consensus of all and the exact location is open to question..

To summarize HTSC in the cuprates begins at high doping with a quantum critical point where the Mott- Hubbard bands separate and orders interact, thereby created a complex suitable for Feshbach resonance pairing.

**5. Conclusions and Suggestions for Future Work.**

We have made arguments supporting the 2e DW experimentally found as an extended "preformed pair" first discussed by Nozieres and Schmitt-Rink. Our argument is strengthened by the experimental and theoretical support for the existence of a similar 2e DW in fullerides such that we propose a general method of pair formation for HTSC involving the Feshbach resonance (FRSC) present in the original BFM and also the MBFM. The unusual DW is generated at a quantum critical point in the phase diagram caused by a singularity in the two body channel. The 2e DW suppresses the BCS-type of superconductivity as doping is reduced past the quantum critical point where the 2e DW also modulates the FRSC, and comprises a component of the Feshbach resonance pairing and then remains as the pseudogap phase as the material passes in an insulating state. Thus the enigmatic pseudogap state is a Peierls-Frolich circular 2e density wave that has a superconductor-like energy gap quasiparticle spectrum, and density of states, but no uniform coherence from one density wave to the next. The present theory seems to account for many STM and other experimental observations in the complete phase diagram. A number of the unusual features have been observed in cold atom work, and it would be interest to know if they can be seen in the solid state. Also, 1t might be interesting to review organic superconductivity in light of our theory.

As this manuscript was nearing completion, the work of T. Yoshioka et al [86] appeared. These authors study the metal-insulator transition in the Hubbard model on the checkerboard lattice. At half-filling, they investigated the effect of electrons correlations in the Hubbard model on this lattice at absolute zero by means of the path-integral renormalizations group method. They demonstrate that the systems exhibit a simple first-order phase transition to the plaquette-singlet phase at a finite value of the Hubbard interaction.

The discovery of the oxypnictide superconductor $LaFeAsO_{1-x}F_x$ (denoted by LFAO-F) with a critical temperature [87] is noteworthy. This class of compounds has a layered structure like high-$T_c$ cuprates where the dopant and conducting layers are separated in such a manner that the doped carriers, namely electrons introduced by the substitution of $O^{2-}$ with $F^-$ in the $La_2O_2$ layers, move within the layers formed by strongly bonded Fe and As atoms. Another gross similarity to the cuprates is that superconductivity occurs upon doping of parent phases exhibiting long range magnetic order. Measurements of muon spin rotation/relaxation ($\mu SR$) on various oxypnictide superconductors indicate that the superfluid density $n_s$ lies on the empirical $n_s$ vs. $T_c$ line observed for the underdoped cuprates [88]. This leads one to suppose that there is a similar mechanism for superconductivity between the high-$T_c$ cuprates and the oxypnictides.

## Appendix A. Mean field Theory for Peierls-Frolich Charge Density Wave Transition.

Treating our two electrons as a 1D free electron gas with a Hamiltonian in second quantized form

$$H = \sum_k \varepsilon_k a_k^\dagger a_k$$

and energy $\varepsilon_k = \hbar^2 k^2 / 2m$ and $a_k^\dagger, a_k$ being the creation and annihilation operators, respectively; spin degrees of freedom are omitted. Following O'Brien [31a], we use an average "collective coordinate" $H_g$ coordinate for all eight vibrations of this symmetry described by Hamiltonian

$$H_{ph} = \sum_q \left\{ \frac{P_q P_{-q}}{2M} + \frac{M\omega_q^2}{2} Q_q Q_{-q} \right\} \quad (A1)$$

where M is the mass, $\omega_q$ are the normal coordinates, and $Q_q$, $P_q$ are the standard normal coordinates and conjugate momenta of the atom motions. This Hamiltonian can be rewritten as

$$H_{ph} = \sum_q \hbar\omega_q \left( b_q^\dagger b_q + \frac{1}{2} \right)$$

using the following

$$Q_q = \left( \frac{\hbar}{2M\omega_q} \right)^{1/2} \left( b_q + b_{-q}^\dagger \right)$$

$$P_q = \left( \frac{\hbar M\omega_q}{2} \right)^{1/2} \left( b_q^\dagger - b_{-q} \right)$$

In this notation the lattice displacement is

$$u(x) = \sum_q \left( \frac{\hbar}{2NM\omega_q} \right)^{1/2} \left( b_q + b_{-q}^\dagger \right) e^{iqx}$$

with $\left( N = lattice-sites/length \right)$. Using the "rigid ion approximation" to describe the electron-vibration interaction assumes that the potential V depends only on the distance from the equilibrium lattice position resulting in

$$H_{el-vib} = \sum_{k,k',l} \langle k | V(r-l-u) | k' \rangle a_k^\dagger a_k$$

$$= \sum_{k,k',l} e^{i(k-k')(l+u)} V_{k-k'} a_k^\dagger a_k$$

Here l is the equilibrium atom position, u is the distance from equilibrium, and $V_{k-k'}$ is the Fourier transform of a single atom potential. Approximating for small displacements

$$e^{i(k'-k)u} \approx 1 + i(k'-k)u = 1 + iN^{-1/2}(k'-k)\sum_q e^{iql} u_q$$

Ignoring the interaction of the electrons with ions in their equilibrium positions, we have

$$H_{el-vib} = iN^{-1/2} \sum_{k,k',l,q} e^{i(k'-k+q)l} (k'-k) u_q V_{k-k'} a_k^\dagger a_k$$

$$= iN^{-1/2} \sum_{k,k'} (k'-k) u_q V_{k-k'} a_k^\dagger a_k$$

Expressing the interaction in second quantization terms,

$$H_{el=vib} = i\sum_{k,k'} \left(\frac{\hbar}{2M\omega_{k-k'}}\right)^{1/2} (k'-k) V_{k-k'} \left(b_{k'-k}^\dagger + b_{k-k'}\right) a_k^\dagger a_{k'}$$

$$= \sum_{k,q} g_q \left(b_{-q}^\dagger + b_q\right) a_{k+q}^\dagger a_k$$

where g, the coupling constant, is

$$g_q = i\left(\frac{\hbar}{2M\omega_q}\right)^{1/2} |q| V_q$$

Putting all three terms together fives what is known as the Fröhlich Hamiltonian

$$H = \sum_k \varepsilon_k a_k^\dagger + \sum_q \hbar\omega b_q^\dagger b_q + \sum_{k,q} g_q a_{k+q}^\dagger a_k \left(b_{-q}^\dagger + b_q\right) \tag{A2}$$

The effect on the normal vibration coordinates for small amplitude displacement is ($\ddot{Q}_q$ is the second derivative with respect to time, and using the equation)

$$\hbar^2 \ddot{Q}_q = -\left[\left[Q_q, H\right], H\right]$$

Since $\left[Q_q, P_{q'}\right] = i\hbar \delta_{q,q'}$, we have

$$\ddot{Q}_q = -\omega_q^2 Q_q - g\left(\frac{2\omega_q}{M\hbar}\right)^{1/2} \rho_q$$

with $\rho_q = \sum_k a_{k+q}^\dagger a_k$ being the $q^{th}$ component of the electron density (g is assumed independent of q). The second term on the RHS is an effective force constant due to combined electron-vibration interaction. The ionic potential $g\left(2M\omega_q/\hbar\right)^{1/2} Q_q$ results in a density fluctuation

$$\rho_q = \chi(q,T) g \left(\frac{2M\omega_q}{\hbar}\right)^{1/2} Q_q$$

and using linear response theory leads to the following equation of motion

$$\ddot{Q}_q = -\left[\omega_q^2 + \frac{2g^2 \omega_q}{M\hbar} \chi(q,T)\right] Q_q$$

which produces a renormalized vibration frequency

$$\omega_{ren,q}^2 = \omega_q^2 + \frac{2g^2 \omega_g}{M\hbar}$$

For our 1D model $\chi(q,T)$ has its maximum value at $q = 2k_F$, the so-called Kohn anomaly where the reduction (softening) of the vibrational frequency will be most significant. Here

$$\omega_{ren,2k_F}^2 = \omega_{2k_F}^2 - \frac{2g^2 n(\varepsilon_F)\omega_{2k_F}}{\hbar}\ln\left(\frac{1.14\varepsilon_0}{k_B T}\right)$$

As the temperature is reduced, the renormalized vibration frequency goes to zero which defines a transition temperature where a frozen-in distortion occurs. From the equation above a mean field transition temperature for a charge density wave, $T_{CDW}^{MF}$, can be calculated

$$k_B T_{CDW}^{MF} = 1.14\varepsilon_0 e^{-1/\lambda} \tag{A3}$$

with $\lambda$, the electron-vibration coupling constant (dimensionless). This is exactly like the BCS equation; the resulting interaction between CDW and BCS SC, discussed in Section 2.

Below the phase transition we are proposing that the $H_g$ collective renormalized vibration frequency is zero since the lattice distortion is "frozen" as a molecular vibration mode with expectation values $\langle b_{2k_F}\rangle = \langle b_{-2k_F}^\dagger\rangle \neq 0$. Defining an order parameter

$$|\Delta|e^{i\phi} = g\left(\langle b_{2k_F}\rangle + \langle b_{-2k_F}^\dagger\rangle\right) \tag{A4a}$$

The lattice displacement can now be shown to be

$$\langle u(x)\rangle = \left(\frac{\hbar}{2NM\omega_{2k_F}}\right)^{1/2}\left\{i\left(\langle b_{2k_F}\rangle + \langle b_{-2k_F}^\dagger\rangle\right)e^{i2k_F} + cc\right\} \tag{A4b}$$

$$= \left(\frac{\hbar}{2NM\omega_{2k_F}}\right)^{1/2}\frac{2\Delta}{g}\cos(2k_F x + \phi) \tag{A4c}$$

$$= \Delta u \cos(2k_F x + \phi) \tag{A4d}$$

along with

$$\Delta u = \left(\frac{2\hbar}{NM\omega_{2k_F}}\right)^{1/2}\frac{|\Delta|}{g}$$

The Fröhlich Hamiltonian gets modified to

$$H = \sum_k \varepsilon_k a_k^\dagger + \sum_q \hbar\omega b_q^\dagger b_q + \sum_{k,q} g_q a_{k+q}^\dagger a_k \langle b_{-q}^\dagger + b_q\rangle \tag{A5}$$

and since $q = \pm 2k_F$ and $\langle b_{2k_F}\rangle = \langle b_{-2k_F}^\dagger\rangle$

$$H = \sum_k \varepsilon_k a_k^\dagger a_k + 2g\sum_k\left[a_{k+2k_F}^\dagger a_k\langle b_{-2k_F}^\dagger\rangle + a_{k-2k_F}^\dagger a_k\langle b_{-2k_F}\rangle\right] + 2\hbar\omega_{2k_F}\langle b_{2k_F}\rangle^2$$

The order parameter defined earlier (eq 18a) is now

$$H_{el} = \sum_k\left[\varepsilon_k a_k^\dagger a_k + |\Delta|e^{i\phi}a_{k+2k_F}^\dagger a_k + |\Delta|e^{-i\phi}a_{k-2k_F}^\dagger a_k\right]$$

The standard approximation is to consider only states near the Fermi level. Labeling states near $+k_F$ by index 1 and those near $-k_F$ by index 2, and using the linear dispersion relationship $\varepsilon_k = \hbar v_F (k - k_F)$,

$$H = \sum_k \left[ \varepsilon_k \left( a_{1,k}^\dagger a_{1,k} - a_{2,k}^\dagger a_{2,k} \right) + |\Delta| e^{i\phi} a_{1,k}^\dagger a_{2,k} + |\Delta| e^{-i\phi} a_{2,k}^\dagger a_{1,k} \right]$$

This Hamiltonian can be diagonalized by a canonical transformation with new operators

$$\gamma_{1,k} = U_k a_{1,k} - V_k^\circ a_{2,k} = U_k e^{-i\phi/2} a_{1,k} - V_k e^{i\phi/2} a_{2,k}$$

and

$$\gamma_{2,k} = V_k a_{1,k} + U_k^\circ a_{2,k} = V_k e^{-i\phi/2} a_{1,k} + U_k e^{i\phi/2} a_{2,k}$$

using the constraint $U_k^2 + V_k^2 = 1$. The Hamiltonian becomes

$$H = \sum_k \left[ \varepsilon_k \left( U_k^2 - V_k^2 \right) - 2|\Delta| U_k V_k \right] \left( \gamma_{1,k}^\dagger \gamma_{1,k} - \gamma_{2,k}^\dagger \gamma_{2,k} \right) + \left[ 2\varepsilon_k U_k V_k + |\Delta| \left( U_k^2 - V_k^2 \right) \right] \left( \gamma_{1,k}^\dagger \gamma_{2,k} + \gamma_{2,k}^\dagger \gamma_{1,k} \right)$$

If the coefficients in front of the off-diagonal terms are zero, the Hamiltonian can be diagonalized

$$\left[ 2\varepsilon_k U_k V_k + |\Delta| \left( U_k^2 - V_k^2 \right) \right] = 0 \quad \text{and} \quad U_k^2 + V_k^2 = 1$$

Using

$$V_k = \cos\left(\frac{\theta_k}{2}\right) \quad \text{and} \quad U_k = \sin\left(\frac{\theta_k}{2}\right)$$

then

$$\tan \theta_k = -\frac{|\Delta|}{\varepsilon}$$

$$V_k^2 = \frac{1}{2}\left( 1 - \frac{\varepsilon_k}{\left( \varepsilon_k^2 + \Delta^2 \right)^{1/2}} \right) = \frac{1}{2}\left( 1 + \frac{\varepsilon_k}{E_k} \right)$$

and

$$U_k^2 = \frac{1}{2}\left( 1 - \frac{\varepsilon_k}{E_k} \right)$$

where

$$E_k - \varepsilon_F + sign(k - k_F)\left[ \hbar^2 v_F^2 (k - k_F)^2 + \Delta^2 \right]^{1/2}$$

Substituting the expressions for $U_k, V_k$ into the Hamiltonian results in

$$H = \sum_k E_k \left( \gamma_{1,k}^\dagger \gamma_{1,k} + \gamma_{2,k}^\dagger \gamma_{2,k} \right) + \frac{\hbar \omega_{2k_F} \Delta^2}{2g^2} \tag{A6}$$

The result is that the linear dispersion $\varepsilon_k - \varepsilon_F = \hbar v_F (k - k_F)$ is no longer valid as a gap is developed in the dispersion. Using a approximation for the density of states

$$N_{CDW}(E) dE = N_e(\varepsilon) d\varepsilon = N_e d\varepsilon$$

$$\frac{N_{CDW}(E)}{N_e} = \frac{d\varepsilon}{dE} = \begin{cases} 0 \\ \dfrac{E}{\left(E^2 - \Delta^2\right)^{1/2}} \end{cases}$$

The first condition applies if $|E| < \Delta$ while the second when $|E| > \Delta$. The **opening of the gap leads to a lowering of electronic energy**

$$E_{el} = \sum_k \left(-E_k + v_k k\right) = n(\varepsilon_F) \int_0^{\varepsilon_F} \left(\varepsilon - \left(\varepsilon^2 + \Delta^2\right)^{1/2}\right) d\varepsilon$$

It can be shown that there are two terms leading to this lowering of energy: 1) an electronic term

$$E_{el} = n(\varepsilon_F)\left[-\frac{\Delta^2}{2} - \Delta^2 \log\left(\frac{2\varepsilon_F}{\Delta}\right)\right] + ...$$

and 2) a lattice term

$$E_{latt} = \frac{N}{2} M \omega_{2k_F}^2 \langle u(x)\rangle^2 = \frac{\hbar \omega_{2k_F} \Delta^2}{2g^2} = \frac{\Delta^2 n(\varepsilon_F)}{\lambda}$$

where $\lambda$ is defined earlier. The total energy change then is

$$E = E_{el} + E_{latt} = n(\varepsilon_F)\left[-\frac{\Delta^2}{2} - \Delta^2 \log\left(\frac{2\varepsilon_F}{\Delta}\right) + \frac{\Delta^2}{2\lambda}\right]$$

For $\lambda \ll 1$ and minimizing the total energy gives

$$\Delta = 2\varepsilon_F e^{-1/\lambda} \qquad (A7)$$

and a condensation energy of

$$E_{cond} = E_{norm} - E_{CDW} = \frac{n(\varepsilon_F)}{2}\Delta^2 \qquad (A8)$$

We propose that the pseudo 1D nature lowers the potential for the two electrons making the preformed pair binding considerably larger and stronger than a "traditional" Cooper pair.

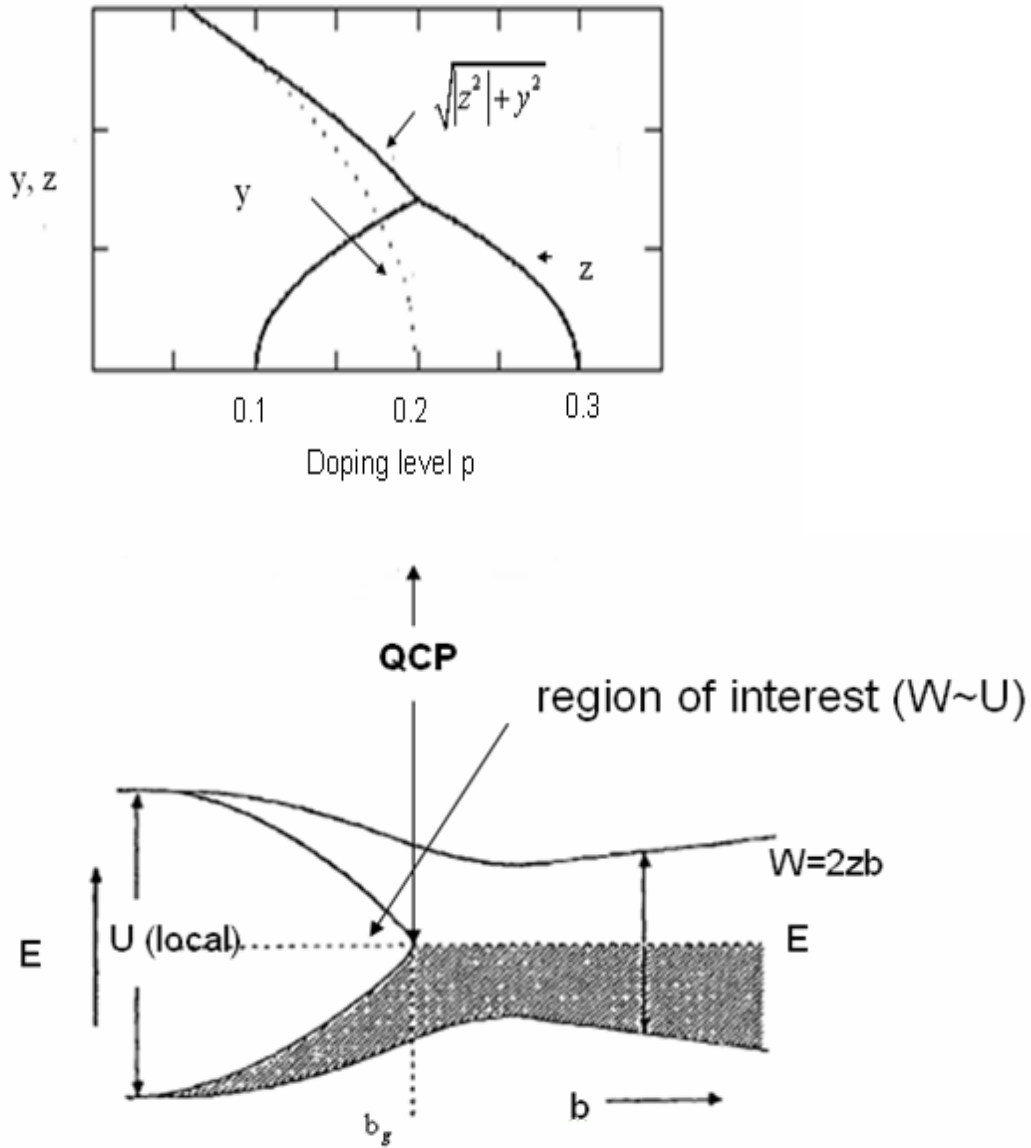

Figure 1. Top (a) is an overview of the order interaction, after [38]. The resulting interaction is expressed in terms of the Ginsburg-Landau free energy, eq (17)
$F = \left(y^2 + |z|^2\right)^2 + \gamma y^2 |z|^2 - \alpha y^2 - \alpha' |z|^2$ where $\alpha, \alpha'$ are adjustable parameters. Bottom (b) Mott-Hubbard model illustration where b is a continuous variable from the insulator to the Fermi liquid metal through the M-I transition. The value of the initial splitting of the bands for both the Cuprates and fullerides is estimated to be about 2 eV [52]

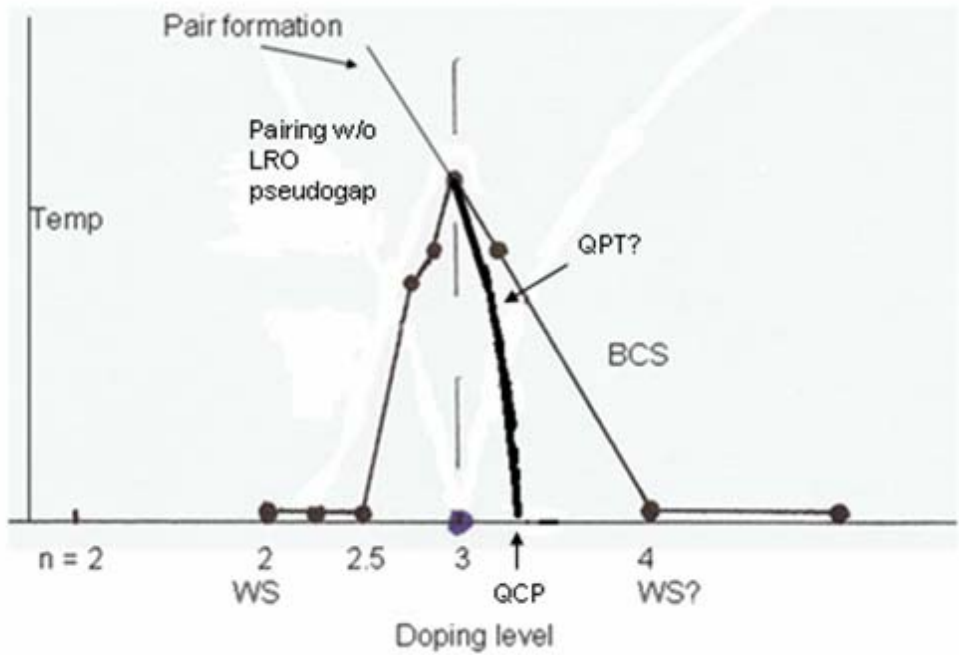

Fig 2a. Revised fulleride phase diagram. Material is superconducting from doping levels of 2.5 to 4 electrons. Preformed pairs are the result of the formation of a 2e density wave (see Appendix A). The underdoped MBFM region results from Feshbach resonance pairing, while the BCS regime is the result of the non-resonant portion (U< 0) from the pairing term, $U_{eff} = U + \dfrac{g_r^2}{2\mu - 2\nu}$

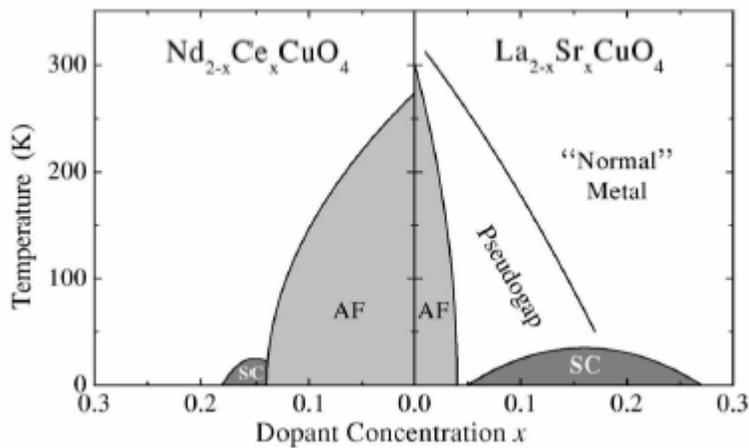

Fig 2b. Proposed phase diagram of electron doped Cuprate superconductors (left side) and hole doped "high temperature superconductors" (right side). From Damascelli et al [89].

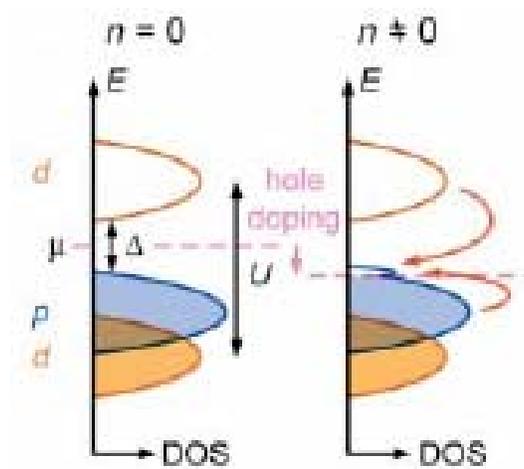

Figure 3. On hole doping a Mott-Hubbard insulator in the localized limit, a two hole state is created [52b].

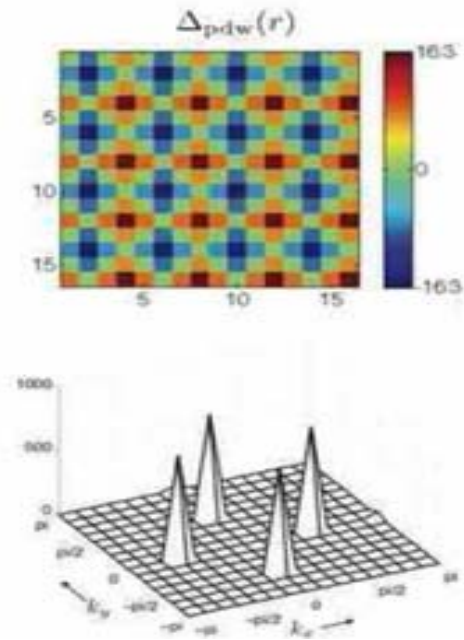

Figure 4. Illustration of the amplitude of the order parameters of the pair density wave (top) and the Fourier transform of the order parameters (bottom) in the first Brillouin zone. The peaks are located at $\vec{Q} = \left[ \left( \pm \pi/2, 0 \right), \left( 0, \pm \pi/2 \right) \right]$ (from [73]).

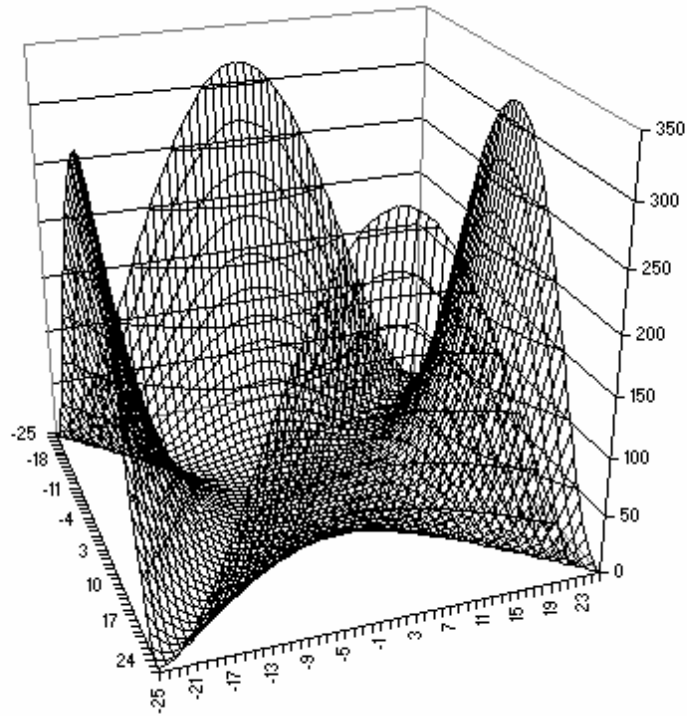

Figure 5. Plot of the approximated Cooper pair function, eq (22),
$\rho = |\psi(\vec{r})|^2 \propto [\cos(rk_F)(x^2 - y^2)e^{-3r/\xi_0}]^2$ with $\xi_0 = 4a$ [77].

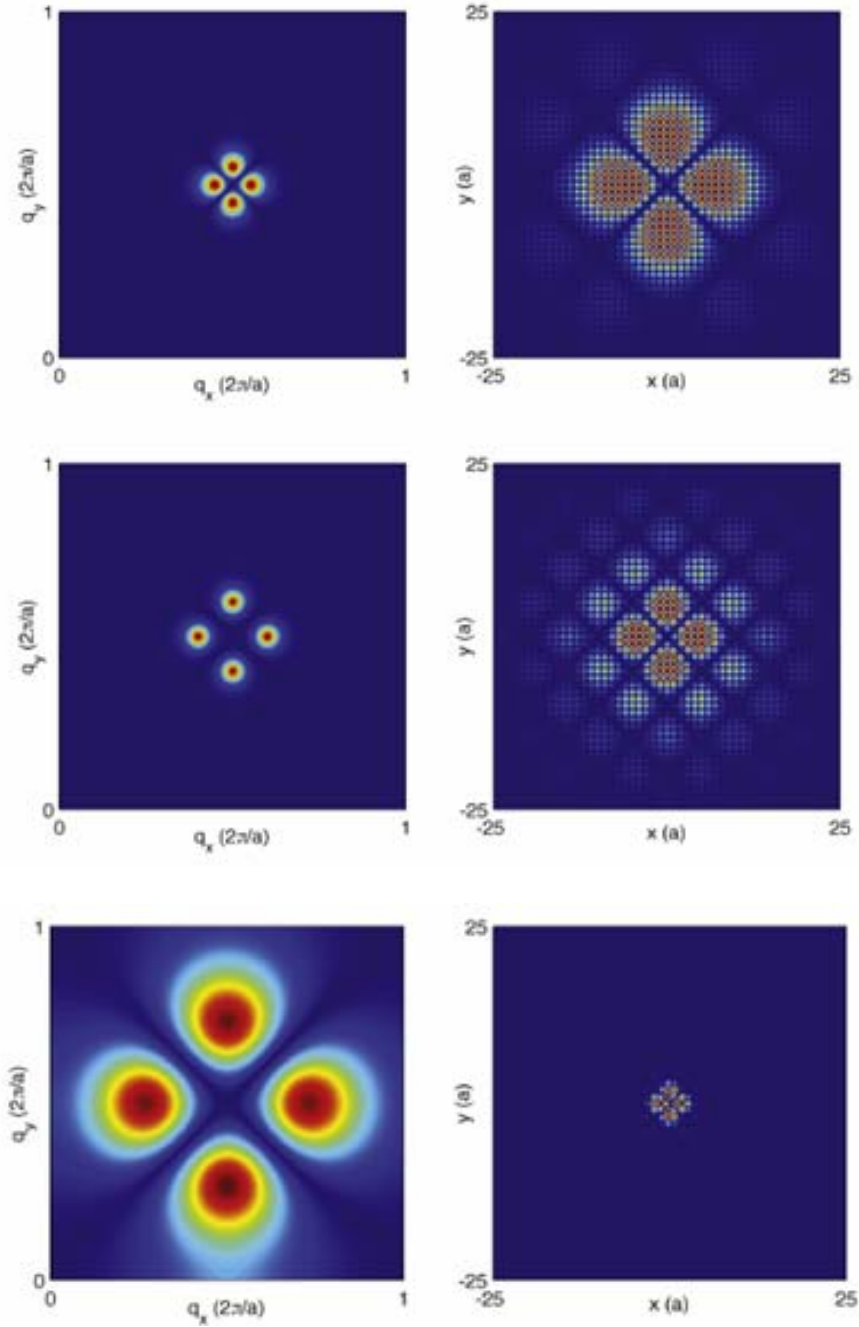

Figure 6 (from [78]). Column one is the incommensurate spin fluctuations (doping levels p equal 0.06, 0.1, and 0.152 (top to bottom) for all illustrations. Column two shows the spatially varying moment obtained by Fourier transform eq (23). A comparison of the moment (column 2) to $\left|\Psi(\vec{r})_{CP}\right|^2$ (Fig 5) is instructive (also see Figure 7).

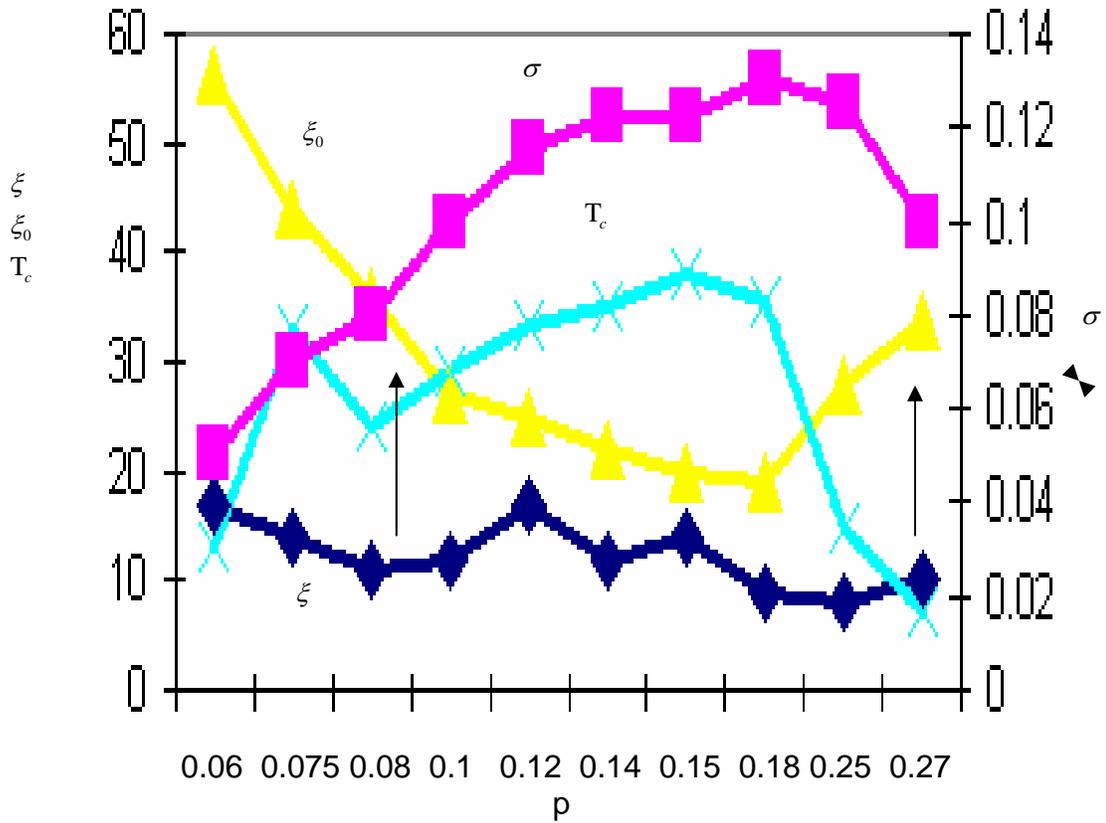

Figure 7. Concurrent graphs of $\xi$, spin correlation length, $\xi_0$, Cooper pair correlation length, $T_c$, and $\delta$, extrapolated values of the incommensurability, from [ ]. As we suggest in the manuscript, resonance is lost at low and high doping (arrows), through for different reasons.